\documentclass[galaxies,article,accept,oneauthor,pdftex,10pt,a4paper]{mdpi}
\usepackage{bm}
\usepackage{amssymb}

\firstpage{1}
\articlenumber{100}
\doinum{10.3390/galaxies5040100}
\pubvolume{5}
\pubyear{2017}
\copyrightyear{2017}
\history{Received: 31 October 2017; Accepted: 1 December 2017; Published: 12 December 2017}
\pdfoutput=1

\Title{Suborbital Fermi/LAT Analysis of the Brightest Gamma-Ray Flare of Blazar 3C~454.3}

\Author{Krzysztof Nalewajko \orcidA{}}
\AuthorNames{Krzysztof Nalewajko}

\address[1]{Nicolaus Copernicus Astronomical Center, Polish Academy of Sciences, Bartycka 18, 00-716 Warsaw, Poland; knalew@camk.edu.pl}

\corres{Correspondence: knalew@camk.edu.pl\textcolor{red}{(Please carefully check the accuracy of names and affiliations. Changes will not be possible after proofreading.)}}

\abstract{
Recent detection of suborbital gamma-ray variability of Flat Spectrum Radio Quasar (FSRQ) 3C~279 by Fermi Large Area Telescope (LAT) is in severe conflict with established models of blazar emission.
This paper presents the results of suborbital analysis of the Fermi/LAT data for the brightest gamma-ray flare of another FSRQ blazar 3C~454.3 in November 2010 (Modified Julian Date; MJD~55516-22).
Gamma-ray light curves are calculated for characteristic time bin lengths as short as 3~min.
The measured variations of the 0.1--10~GeV photon flux are tested against the hypothesis of steady intraorbit flux.
In addition, the structure function is calculated for absolute photon flux differences and for their significances.
Significant gamma-ray flux variations are measured only over time scales longer than $\sim$5 h, which is consistent with the standard blazar models.
}


\begin{document}

\section{Introduction}

Blazars---active galaxies with relativistically beamed non-thermal broad-band emission---belong to the brightest gamma-ray sources in the sky. Their emission is notoriously variable, and their gamma-ray variability has been measured on time scales ranging from years to minutes (see \cite{mad16} for a recent review). The Fermi Large Area Telescope is the most sensitive instrument for measuring high-energy (HE) gamma rays in the energy range 0.1--10 GeV \cite{atw09}. Since August 2008, it performs an almost uninterrupted monitoring of the entire sky, orbiting Earth at the period of $95\;{\rm min}$.

Determining the shortest variability time scale for the gamma-ray emission of blazars has tremendous theoretical implications for the physics of energy dissipation and particle acceleration in relativistic jets. Before the launch of Fermi, the shortest variability time scales of $t_{\rm var} \sim 2\;{\rm min}$ were measured in the very-high-energy (VHE) gamma rays above $100\;{\rm GeV}$ by ground-based Cherenkov telescopes, in particular by H.E.S.S. in PKS~2155-304 \cite{aha07} and by MAGIC in Mrk~501~
\cite{alb07}. Such a short variability time scale is much shorter than the light-crossing time $t_{\rm bh} \sim 3 M_{\rm bh,9}\;{\rm h}$ of supermassive black holes located at the bases of relativistic jets of blazars. This implicates the existence of very compact local dissipation sites, e.g., related to relativistic magnetic reconnection \cite{gia09, nal11}, or extremely efficient jet focusing due to recollimation shocks \cite{bod17}. In addition, the combination of high apparent luminosity with small size of the emitting region, i.e., exceptional radiative compactness, poses a problem of potentially very efficient intrinsic absorption of gamma-ray photons. Additional challenges arise in the case of the most luminous blazars known as flat spectrum radio quasars (FSRQs) due to the presence of dense radiation fields that provide a target for \emph{external absorption} of gamma rays~\cite{tav11,nal12}. In order to avoid such absorption, one can consider highly relativistic bulk motions with Lorentz factors $\Gamma \sim 100$~\cite{beg08} or possibly conversion of gamma-ray photons into axions \cite{tav12}.

There were previously claims of the detection of gamma-ray variability in Fermi/LAT sources at suborbital time scales, e.g., in PKS~1510-089 \cite{fos13}. When analysing the brightest gamma-ray flares of blazars during the first four years of the Fermi mission \cite{nal13}, the very brightest case of 3C~454.3 at MJD~55520 (November 2010) \cite{abd11} was investigated for possible evidence of suborbital variability. That~analysis, performed with the older calibration standard {\tt P7V6}, was inconclusive, and hence it was not published at that time (upper limits on variability time scales of a few hours were reported for this event by~\cite{fos11}).
However, after the Fermi Collaboration presented the case for suborbital variability in blazar 3C~279 at MJD $\simeq$ 57189 (Jun 2015) \cite{ack16}, it became clear that the case of 3C~454.3 needs to be reconsidered. Indeed, the theoretical implications of detectable suborbital gamma-ray variability in 3C~279 are extreme (see also \cite{pet17,vit17,aha17}).
For example, in the standard ERC (External Radiation Comptonization) scenario of gamma-ray emission in FSRQ blazars, the minimum jet Lorentz factor should be $\Gamma_{\rm min} \simeq 50$ to satisfy the opacity, cooling, Eddington and SSC constraints, and $\Gamma_{\rm eqp} \simeq 120$ to achieve equipartition between matter and magnetic fields \cite{ack16}. Moreover, a large fraction of the total jet power should be concentrated into a tiny emitting region of \mbox{$R_\gamma \simeq 10^{-4}(\Gamma/50)\;{\rm pc}$} at the distance scale comparable to the broad-line region $r_\gamma \gtrsim r_{\rm BLR} \simeq 0.1\;{\rm pc}$.
Such requirements cannot be reconciled with the conventional models of blazar emission \cite{der09,sik09,ghi10,boe13,nal14}.

The results of suborbital Fermi/LAT analysis for 3C~454.3 are presented in Section \ref{sec_res}. A comparison of these results with the case of 3C~279 is discussed in Section \ref{sec_dis}. Details of the analysis method are described in Section \ref{sec_met}. Section \ref{sec_con} presents the conclusions.

\section{Results}
\label{sec_res}

\begin{figure}[H]
\centering
\includegraphics[width=0.99\textwidth]{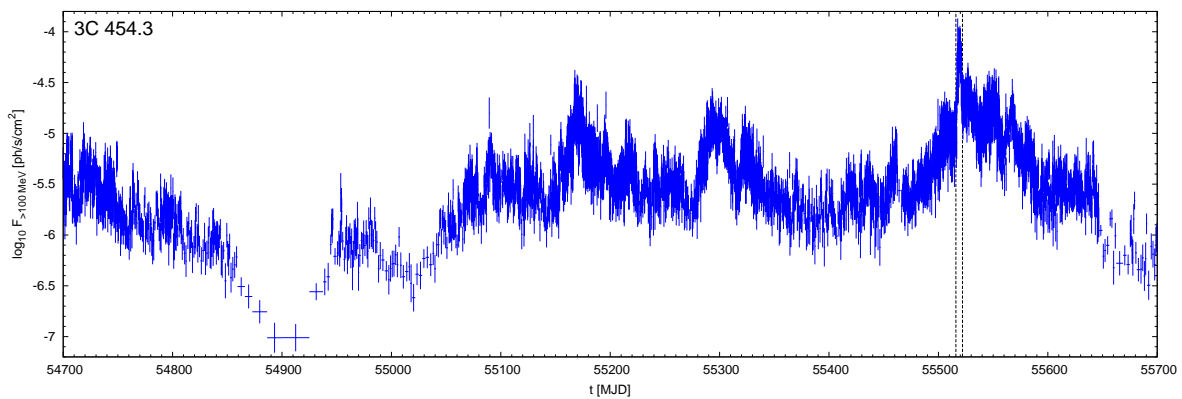}
\caption{Long-term gamma-ray light curve ($E > 100\;{\rm MeV}$) of blazar 3C 454.3 calculated from Fermi Large Area Telescope (LAT) data using adaptive time bins with Test Statistic (TS) $> 25$
\cite{sob14}. The dashed vertical lines indicate the time range covered by Figure \ref{fig_lc_med}.}
\label{fig_lc_long}
\end{figure}

Figure \ref{fig_lc_long} shows the long-term gamma-ray variations of blazar 3C~454.3, calculated with the use of adaptive time bins \cite{sob14}. Over the period of 1000 days, the gamma-ray flux of 3C~454.3 varies by almost factor 1000. Correspondingly, the lengths of the time bins range from 25 days down to $\sim$20~min, always satisfying the detection condition of ${\rm TS} > 25$. During this time, 3C~454.3 produced three major outbursts. The first two, peaking at MJD~55167 and MJD~55294, were originally investigated in \cite{ack10}. The~last one, peaking at MJD~55520 with $F_{>100\;{\rm MeV}} \sim 8.5\times 10^{-5}\;{\rm ph\,s^{-1}\,cm^{-2}}$ (for 3 h bins)~\cite{abd11}, represents the brightest gamma-ray state of any blazar to date, also exceeding the brightest gamma-ray~pulsars.

\vspace{-6pt}

\begin{figure}[H]
\centering
\includegraphics[width=\textwidth]{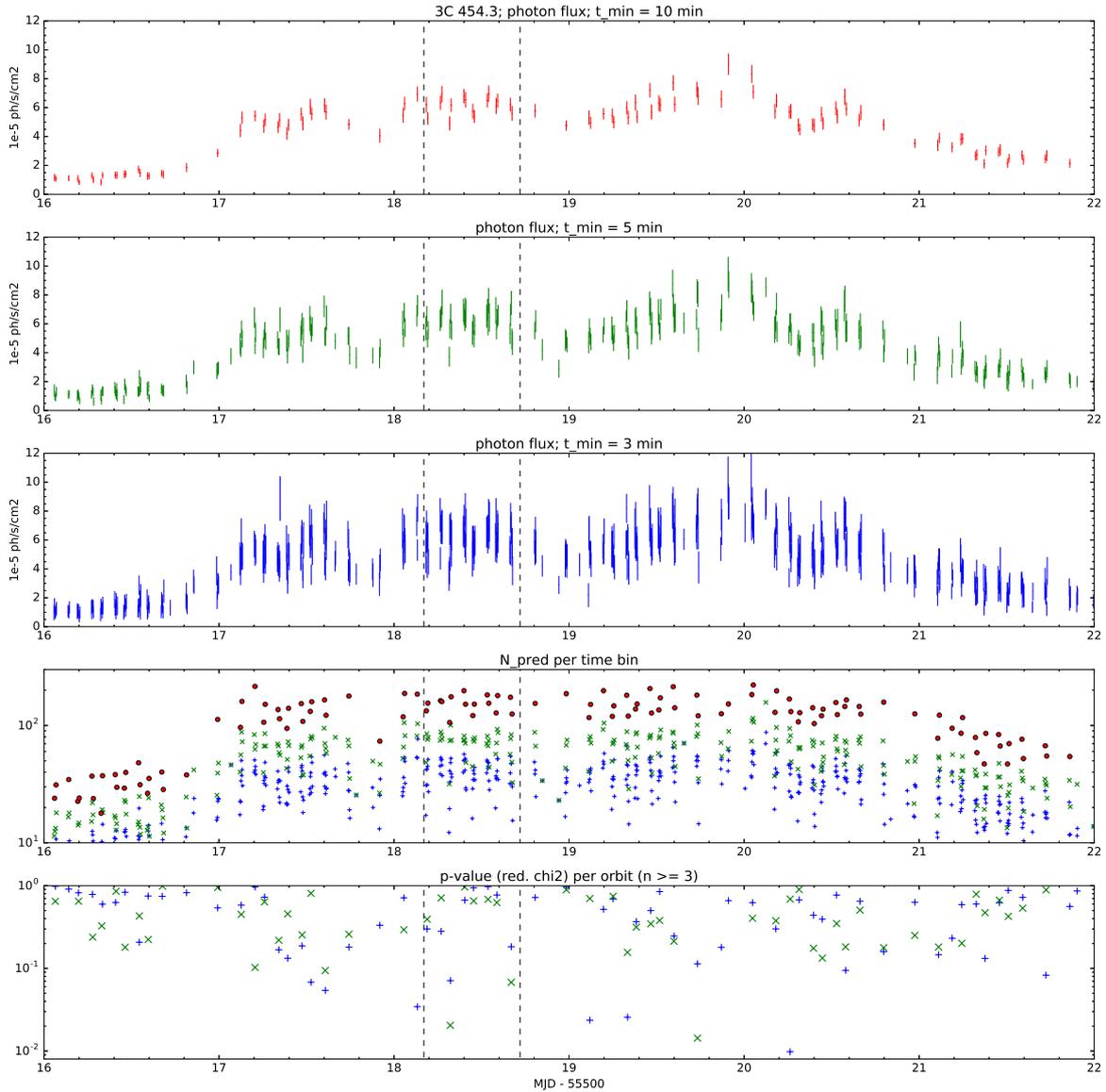}
\caption{{\bf Top three panels:} gamma-ray light curves ($E > 100\;{\rm MeV}$) of blazars 3C 454.3 extracted from the Fermi/LAT data during the brightest gamma-ray flare in Nov 2010 for three values of $t_{\rm min}$.
{\bf Fourth~panel:} the values of photon counts $N_{\rm pred}$ per time bin.
{\bf Bottom panel:} the values of probability $p$ that photon flux variations measured for individual orbits (with at least three independent time bins) are consistent with the null hypothesis of constant flux.
The dashed vertical lines indicate the time range covered by Figure \ref{fig_lc_short}.}
\label{fig_lc_med}
\end{figure}

Figure \ref{fig_lc_med} presents the gamma-ray light curves of 3C~454.3 produced on sub-orbital time scales in the time range MJD~55516-22. These light curves are calculated for three values of characteristic minimum time bin length $t_{\rm min} = 10, 5, 3\;{\rm min}$. As the lengths of visibility windows appear modulated on a superorbital time scale, the number of time bins per orbit is variable.

The maximum likelihood analysis returns the predicted number $N_{\rm pred}$ of gamma-ray photons (events/counts) associated with 3C~454.3. It is a good measure of the relative measurement error $\delta F/F \simeq N_{\rm pred}^{-1/2}$, or the signal-to-noise ratio ${\rm SNR} \simeq N_{\rm pred}^{1/2}$. Within the time range of MJD~55517-21, we find persistent values of $N_{\rm pred}$ per time bin, with the median values $N_{\rm pred,med} = 147, 70, 37$ for $t_{\rm min} = 10, 5, 3\;{\rm min}$, respectively.

For every orbit where we have at least three independent consecutive measurements (there is no such case for $t_{\rm min} = 10\;{\rm min}$), we perform a reduced $\chi^2$ test against the null hypothesis that the photon flux is constant within the orbital visibility window. In Figure \ref{fig_lc_med}, we report the probability values $p$ for the null hypothesis. We find the median values of $p_{\rm med} = 0.40, 0.58$ for $t_{\rm min} = 5, 3\;{\rm min}$, respectively. The~smallest values that we found are $\sim$$10^{-2}$; hence, we can hardly reject the null hypothesis and claim statistically significant suborbital flux variability.

\vspace{-6pt}

\begin{figure}[H]
\centering
\includegraphics[width=\textwidth]{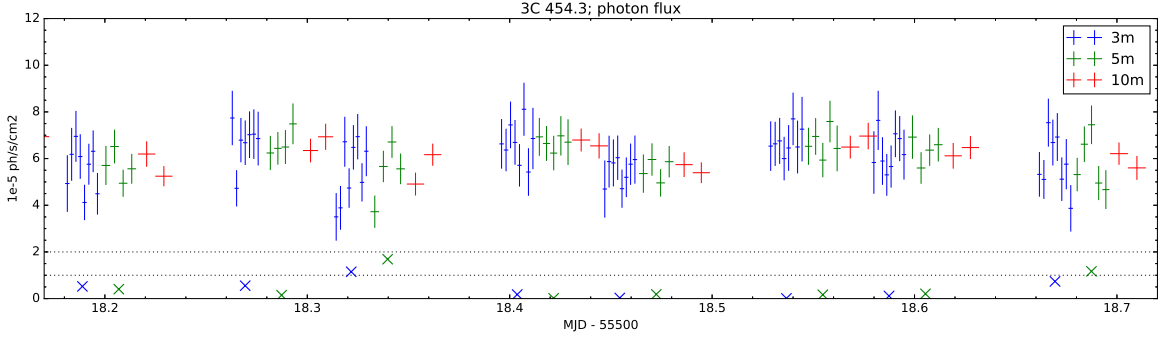}
\caption{A section of light curves shown in Figure \ref{fig_lc_med}. The observation time for the cases of \mbox{$t_{\rm min} = 5,10\;{\rm min}$} (green, red) are shifted by $+0.018,+0.036\;{\rm d}$ with respect to the case of \mbox{$t_{\rm min} = 3\;{\rm min}$} (blue).
 The `x' marks indicate the value of $-\log_{10}p$, where $p$ is the probability of steady intraorbit~flux.}
\label{fig_lc_short}
\end{figure}

Figure \ref{fig_lc_short} shows a short section (MJD~55518.17-72) of the suborbital light curves, comparing directly the flux measurements for $t_{\rm min} = 10, 5, 3\;{\rm min}$. Within this time range, the strongest departure from constant flux is found for the orbit centred at MJD~55518.32.
For $t_{\rm min} = 5\;{\rm min}$, we have four measurements with $\chi^2/{\rm dof} = 9.79/3$, which yields $p = 0.02$. We also measure the weighted mean photon flux $F_{\rm mean} = 5.41$ (in units of $10^{-5}\;{\rm ph\,s^{-1}\,cm^{-2}}$), the root-mean-square (rms) of flux ${\rm rms}(F) = 1.07$, and the rms of flux statistical error ${\rm rms}(\delta F) = 0.68$.
For comparison, in the case of $t_{\rm min} = 3\;{\rm min}$, we have eight~measurements with $\chi^2/{\rm dof} = 13.04/7$, $p = 0.07$, $F_{\rm mean} = 5.45$, ${\rm rms}(F) = 1.26$, and ${\rm rms}(\delta F) = 0.97$.


\vspace{-6pt}

\begin{figure}[H]
\centering
\includegraphics[width=\textwidth]{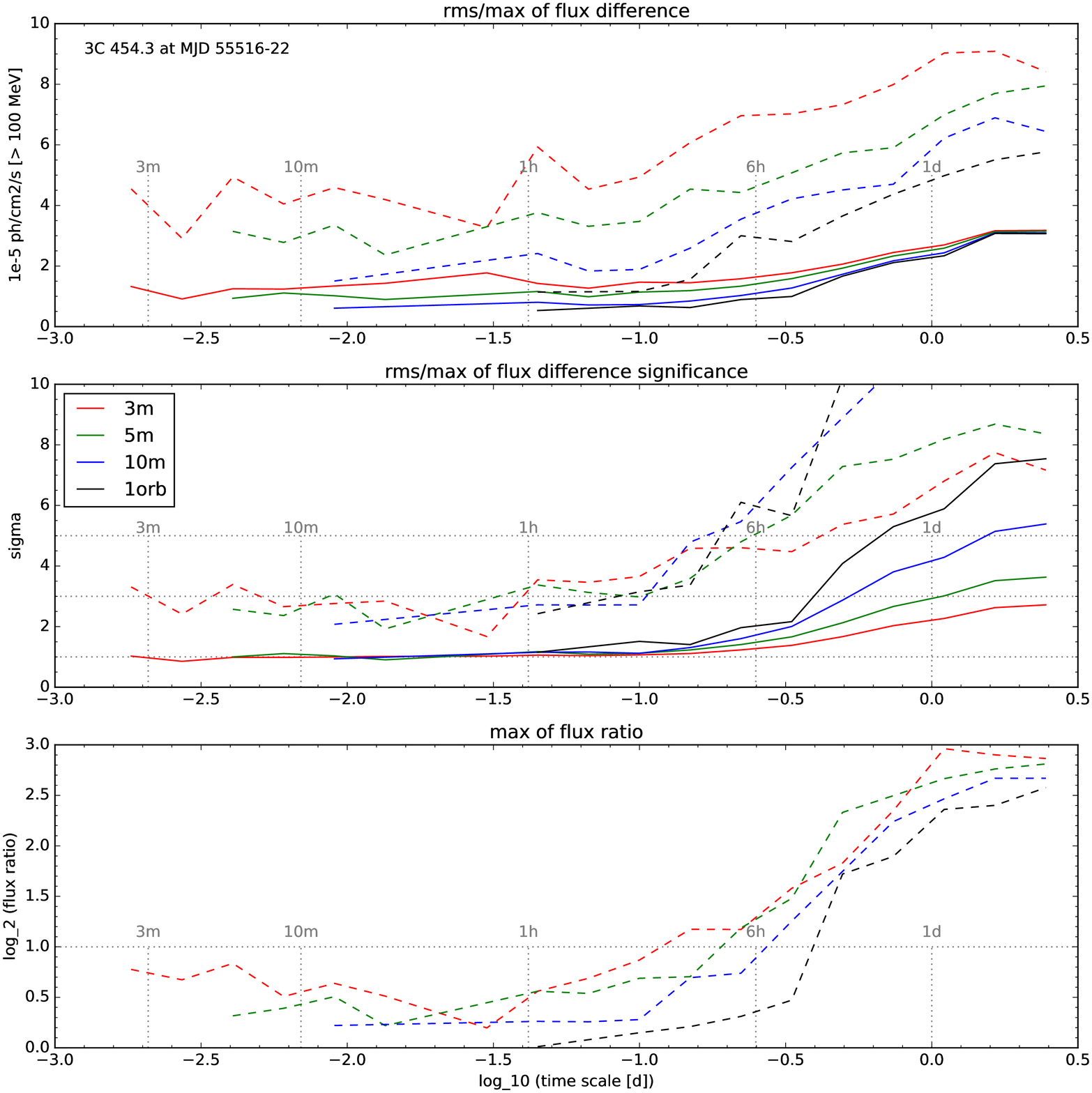}
\caption{Variability statistics as functions of observed time scale for the gamma-ray flux of 3C~454.3 during the major flare at MJD~55516-22.
{\bf Top panel} shows the ${\rm rms}(\Delta f)$ (solid lines) and $\max|\Delta f|$ (dotted lines) for the distributions of photon flux differences $\Delta f = f_2-f_1$.
{\bf Middle panel} shows the ${\rm rms}(\sigma_f)$ (solid lines) and $\max|\sigma_f|$ (dotted lines) for the distributions of photon flux difference significances $\sigma_f = \Delta f/\delta f$.
{\bf Bottom panel} shows the $\max(R_f)$ (dotted lines) for the distributions of photon flux ratios $R_f = f_2/f_1$.
The line colours correspond to different characteristic binning time~scales.
}
\label{fig_stats}
\end{figure}

Figure \ref{fig_stats} shows the results of structure function analysis. As described in Section \ref{sec_met}, structure function is calculated for (1) absolute photon flux difference, (2) statistical significance of photon flux difference, and (3) photon flux ratio.
For a particular light curve, the time delay values $\Delta t$ probe the range from $t_{\rm min}$ to about three days.
We use the root-mean-square (rms) statistic (square root of classical structure function) to probe systematic variations, and the maximum (max) statistic to probe occasional variations (shots).
The rms of absolute flux difference decreases systematically with increasing $t_{\rm min}$ for $\Delta t < 1\;{\rm d}$, as expected for statistical noise. The rms values converge at ${\rm rms}(\Delta f) \simeq 3\times 10^{-5}\;{\rm ph\,s^{-1}\,cm^{-2}}$ for $\Delta t \sim 2\;{\rm d}$. The maximum values decrease systematically and do not converge again as expected.

On the other hand, the rms of statistical significance of flux difference converges for short time delays $\Delta t < 2\;{\rm h}$ at $\sigma_{\Delta f} \simeq 1$. For the same range of time delays, the maximum values of $\sigma_{\Delta f}$ are close to 3.
For longer time delays, the rms of $\sigma_{\Delta f}$ exceeds unity, systematically increasing with increasing $t_{\rm min}$. For example, for $\Delta t = 1\;{\rm d}$, we find that ${\rm rms}(\sigma_{\Delta f}) \simeq 2$ for $t_{\rm min} = 3\;{\rm min}$, and ${\rm rms}(\sigma_{\Delta f}) \simeq 6$ for $t_{\rm min} = 95\;{\rm min}$ (1-orbit light curve). The maximum values of $\sigma_{\Delta f}$ also exceed 3 for $\Delta t > 2\;{\rm h}$. The~significance of occasional flux variations (max) exceeds 5 for $\Delta t > 5\;{\rm h}$ (for $t_{\rm min} = 10\;{\rm min}$ and $1\;{\rm orb}$). However, the significance of systematic flux variations (rms) exceeds 3 only for $\Delta t > 9\;{\rm h}$.

In order to estimate the flux-doubling time scale $\tau_{\rm R2} = \Delta t(R_f = 2)$, the max values of the flux ratio are evaluated. Estimates of the flux-doubling time scale are found to increase systematically with $t_{\rm min}$, from $\tau_{R2} \simeq 3\;{\rm h}$ for $t_{\rm min} = 3\;{\rm min}$, to $\tau_{R2} \simeq 8\;{\rm h}$ for $t_{\rm min} = 1\;{\rm orb}$.

\section{Discussion}
\label{sec_dis}

This analysis of blazar 3C~454.3 at MJD~55520 can be compared with the recent study of blazar 3C~279 at MJD~57189 \cite{ack16}. In the case of 3C~279, the exposure to gamma-rays per orbit was enhanced by factor $\sim 3$ thanks to first successful pointing observation by Fermi/LAT. Although occultations by the Earth were still significant, the visibility windows were longer and the exposure to the target more uniform. Nevertheless, the number of gamma-ray counts $N_{\rm pred}$ collected over very short time bins, $t_{\rm bin} \sim 3\;{\rm min}$, is on average higher in the case of 3C~454.3, since its average photon flux is higher by factor $\simeq 2$. Moreover, such high gamma-ray flux was sustained over a longer period of almost four days, allowing for many more suborbital detections.

At redshift $z = 0.859$, the luminosity distance to 3C~454.3 is $d_L \simeq 5.55\;{\rm Gpc}$, hence \mbox{$4\pi d_{\rm L}^2 \simeq 3.7\times 10^{57}\;{\rm cm^2}$}. An upper limit on suborbital photon flux variation amplitude of \mbox{$F_{\rm suborb} < 6\times 10^{-6}\;{\rm ph\,s^{-1}\,cm^{-2}}$} (the rms of photon flux difference for $t_{\rm bin} = 10\;{\rm min}$ at $\Delta t \simeq 12\;{\rm min}$) with the mean photon energy of $E_{\rm mean} \simeq 0.37\;{\rm GeV}$ (corresponding to the photon index of $\Gamma = 2.181$; see below) can be translated into the apparent gamma-ray luminosity of $L_{\gamma,\rm suborb} < 1.3\times 10^{49}\;{\rm erg\,s^{-1}}$. This limit is comparable with the apparent luminosity of suborbital variations detected in blazar 3C~279 ($F_{\rm suborb} \simeq 10^{-5}\;{\rm ph\,s^{-1}\,cm^{-2}}$, $d_{\rm L} \simeq 3.11\;{\rm Gpc}$) \cite{ack16}. Because of the high redshift of 3C~454.3, the eventual detection of gamma-ray variability on the time scales of several minutes would create even more serious problems to the theory of blazars.


\section{Materials and Methods}
\label{sec_met}

The analysis of Fermi/LAT data presented in this work is performed with the final software package {\tt Science Tools} version {\tt v10r0p5}\footnote{\url{https://fermi.gsfc.nasa.gov/ssc/data/analysis/software/}} and with the final instrument calibration standard {\tt P8R2\_SOURCE\_V6}. Photons are selected in the energy range 0.1--10 GeV from a Region of Interest (RoI) of $10^\circ$, applying a zenith angle cut of $<100^\circ$. Background sources were selected from the 3FGL catalog~\cite{ace15} within the radius of $25^\circ$. We applied a detection criterion ${\rm TS} > 10$ and $N_{\rm pred} > 3$.

Although Fermi/LAT is characterised by very wide field-of-view (2.4 sr), individual sources are visible only during short time intervals for each orbit. Using the spacecraft telemetry data, I select visibility windows when the angular separation between the source and the main axis of LAT is $\alpha_{\rm src} < 60^\circ$, also selecting for good-time-intervals (GTIs) and avoiding the South Atlantic Anomaly (SAA).
Such visibility windows are of variable length, but typically they are shorter than $\sim$30 min.
Given a minimum time scale $t_{\rm min}$, each visibility window of duration $T_{\rm vis}$ is divided into as many as possible time bins of equal length $t_{\rm bin} \ge t_{\rm min}$, i.e., $t_{\rm bin} = T_{\rm vis}/{\rm floor}(T_{\rm vis}/t_{\rm min})$.
While this choice results in slightly different values of $t_{\rm bin}$ for each orbit, it assures that the entire exposure of 3C~454.3 (except windows shorter than $t_{\rm min}$) is used in the analysis.

A standard maximum likelihood analysis is used to measure the gamma-ray flux of 3C~454.3. In~order to minimise the number of degrees of freedom, all other parameters, including the normalisations and photon indices of background sources, were fixed to their average values determined from a global fit performed over time range MJD~55516-22. The photon index of 3C~454.3 was fixed at the value $\Gamma = 2.181$ determined in the same way.

From the maximum likelihood analysis, light curves are obtained in the form $({t_i,f_i,\delta f_i})$, where $t_i$ is the centre of time bin, $f_i$ is the measured photon flux, and $\delta f_i$ is the statistical 1 {$-$} $\sigma$ error of photon flux measurement.
A structure function \cite{emm10} is calculated from a given light curve by considering all pairs of measurements $(t,f,\delta f)$ made at times $t_1 < t_2$, binned according to the logarithm of time delay $\Delta t = t_2-t_1$.
Given a measured parameter $F$, for every delay bin, the distribution of differences $\Delta F$ is determined, and then the statistics ${\rm rms}(\Delta F)$ and $\max(\Delta F)$ are calculated.
In particular, three types of variations are considered:
(1) absolute photon flux difference $\Delta f = |f_2 - f_1|$;
(2) significance of photon flux difference $\sigma_f = |f_2-f_1|/\sqrt{(\delta f_1)^2+(\delta f_2)^2}$; and
(3) photon flux ratio $R_f = f_2/f_1$.

\section{Conclusions}
\label{sec_con}

Analysis of gamma-ray variability of blazar 3C~454.3 during its brightest gamma-ray flare at MJD 55516-22 from the Fermi/LAT data is performed on suborbital time scales ($t_{\rm bin} < 95\;{\rm min}$). The statistical significance of photon flux measurements is certainly not worse than in the case of blazar 3C~279 around MJD 57189 \cite{ack16}, where exposure was increased thanks to a successful pointing observation. By probing different characteristic suborbital time scales $t_{\rm min} = 3,5,10\;{\rm min}$, no evidence is found for statistically significant suborbital variability. The reduced $\chi^2$ test against the null hypothesis of constant flux per orbital visibility window returns $p > 10^{-2}$. The structure function analysis suggests: (1) an upper limit on suborbital variations at $\Delta F < 6\times 10^{-6}\;{\rm ph\,s^{-1}\,cm^{-2}}$ corresponding to $\Delta L < 1.3\times 10^{49}\;{\rm erg\,s^{-1}}$; (2) occasional (max) flux variations become significant ($5\sigma$) for $\tau > 5\;{\rm h}$; (3)~systematic (rms) flux variations become significant ($3\sigma$) for $\tau > 9\;{\rm h}$; and (4) the flux-doubling time scale is of the order $\tau_{R2} \sim 6\;{\rm h}$. These results are consistent with the standard models of blazar emission.

\vspace{6pt}

\acknowledgments{Discussions with Greg Madejski and Alex Markowitz are acknowledged.  This work was supported by the Polish National Science Centre grant 2015/18/E/ST9/00580.
This work is based on public data acquired by the Fermi Large Area Telescope created by NASA and DoE (USA) in collaboration with institutions from France, Italy, Japan and Sweden.}

\conflictsofinterest{The author declares no conflict of interest.}


\reftitle{References}

\end{document}